\documentstyle[twoside,fleqn,espcrc2]{article}

\newcommand{\AmS}{{\protect\the\textfont2
  A\kern-.1667em\lower.5ex\hbox{M}\kern-.125emS}}

\title{Perturbative and 
non-perturbative monodromies in $N=2$ heterotic string vacua}

\author{Gabriel Lopes Cardoso\address{Theory Division, CERN, CH-1211 Geneva 23}
        \thanks{Talk presented by G. L. Cardoso}
         Dieter L\"{u}st 
        and 
        Thomas Mohaupt\address{Institut f\"{u}r Physik, Humboldt 
         Universit\"{a}t Berlin, Invalidenstra\ss e 110, D-10115 Berlin}}

\begin{document}

\def\x{\times}
\def\beq{\begin{equation}}
\def\eeq{\end{equation}}
\def\beqa{\begin{eqnarray}}
\def\eeqa{\end{eqnarray}}
\def\D{ {\cal D}}
\def\L{ {\cal L}}
\def\C{ {\cal C}}
\def\N{ {\cal N}}
\def\calE{{\cal E}}
\def\lin{{\rm lin}}
\def\Tr{{\rm Tr}}
\def\mxth{\mathsurround=0pt }
\def\xversim#1#2{\lower2.pt\vbox{\baselineskip0pt \lineskip-.5pt
x  \ialign{$\mxth#1\hfil##\hfil$\crcr#2\crcr\sim\crcr}}}
\def\simgr{\mathrel{\mathpalette\xversim >}}
\def\simle{\mathrel{\mathpalette\xversim <}}

\def\a{\alpha}
\def\b{\beta}
\def\dota{ {\dot{\alpha}} }
\def\lag{Lagrangian}
\def\Kahler{K\"{a}hler}
\def\kahler{K\"{a}hler}
\def\A{ {\cal A}}
\def\C{ {\cal C}}
\def\D{ {\cal D}}
\def\F{{\cal F}}
\def\L{ {\cal L}}

\def\R{ {\cal R}}
\def\x{ \times }
\def\beq{\begin{equation}}
\def\eeq{\end{equation}}
\def\beqa{\begin{eqnarray}}
\def\eeqa{\end{eqnarray}}

\newcommand{\be}{\begin{equation}}
\newcommand{\eq}{\end{equation}}
\newcommand{\ov}{\overline}
\newcommand{\un}{\underline}
\newcommand{\p}{\partial}
\newcommand{\la}{\langle}
\newcommand{\ra}{\rangle}
\newcommand{\bl}{\boldmath}
\newcommand{\ds}{\displaystyle}
\newcommand{\nl}{\newline}
\newcommand{\Nzahl}{{\bf N}  }
\newcommand{\zzahl}{ {\bf Z} }
\newcommand{\Zzahl}{ {\bf Z} }
\newcommand{\Qzahl}{ {\bf Q}  }
\newcommand{\Rzahl}{ {\bf R} }
\newcommand{\Czahl}{ {\bf C} }
\newcommand{\wt}{\widetilde}
\newcommand{\wh}{\widehat}
\newcommand{\fs}[1]{\mbox{\scriptsize \bf #1}}
\newcommand{\ft}[1]{\mbox{\tiny \bf #1}}
\newtheorem{satz}{Satz}[section]
\newenvironment{Satz}{\begin{satz} \sf}{\end{satz}}
\newtheorem{definition}{Definition}[section]
\newenvironment{Definition}{\begin{definition} \rm}{\end{definition}}
\newtheorem{bem}{Bemerkung}
\newenvironment{Bem}{\begin{bem} \rm}{\end{bem}}
\newtheorem{bsp}{Beispiel}
\newenvironment{Bsp}{\begin{bsp} \rm}{\end{bsp}}




\def\S4{\frac{SO(4,2)}{SO(4) \otimes SO(2)}}
\def\P3{\frac{SO(3,2)}{SO(3) \otimes SO(2)}}
\def\MGd{\frac{SO(r,p)}{SO(r) \otimes SO(p)}}
\def\SOd{\frac{SO(r,2)}{SO(r) \otimes SO(2)}}
\def\SO2{\frac{SO(2,2)}{SO(2) \otimes SO(2)}}
\def\SUm{\frac{SU(n,m)}{SU(n) \otimes SU(m) \otimes U(1)}}
\def\SUS{\frac{SU(n,1)}{SU(n) \otimes U(1)}}
\def\SK{\frac{SU(2,1)}{SU(2) \otimes U(1)}}
\def\SU{\frac{ SU(1,1)}{U(1)}}

\begin{abstract}

In this talk we summarise our recent results on perturbative and 
non-perturbative monodromies in four-dimensional heterotic strings
with $N=2$ space-time supersymmetry, and we compare our results
with the rigid $SU(2)$ $N=2$ Yang-Mills monodromies.

\end{abstract}

\maketitle

\section{Classical results, enhanced gauge symmetries and Weyl
reflections \label{classres}}

We will, in the following,  consider four-dimensional $N=2$
heterotic string vacua which are based on compactifications of
six-dimensional $N=1$ heterotic vacua on a two-torus $T_2$. The moduli of $T_2$
are commonly denoted by $T$ and $U$ where $U$ describes the
deformations of the complex structure, $U=(\sqrt G-iG_{12})/G_{11}$
($G_{ij}$ is the metric of $T_2$), while $T$ parametrizes the
deformations of the area and of the antisymmetric tensor, $T=2(\sqrt G+iB)$.
(Other possibly existing vector fields will not be considered in 
the following.)
The scalar fields $T$ and $U$ are the spin-zero component fields of
two $U(1)$ $N=2$ vector supermultiplets.
Classically, the physical properties of two-torus compactifications are
invariant under the group $SO(2,2,{\bf Z})$ of discrete target space
duality transformations. It contains the $T\leftrightarrow U$
exchange symmetry, with group element denoted by $\sigma$, as well as the
$PSL(2,{\bf Z})_T\times PSL(2,{\bf Z})_U$ dualities, which act on $T$
and $U$ as
\be
(T,U) \longrightarrow \left(
\frac{aT - ib}{icT + d}, \frac{a'U - ib'}{ic'U +d'} \right)
\eq
where $ad-bc=a'd'-b'c'=1$.
The classical monodromy group,
which is a true symmetry of the classical effective Lagrangian,
is generated \cite{AFGNT} by the elements $\sigma$,
$g_1$, $g_1$: $T\rightarrow 1/T$ and $g_2$, $g_2$: $T\rightarrow 1/(T-i)$.
The transformation $t$: $T\rightarrow
T+i$, which is of infinite order, corresponds to $t=g_2^{-1}g_1$.
$PSL(2,{\bf Z})_T$ is generated by $g_1$ and $g_2$,
whereas the corresponding elements in $PSL(2,{\bf Z})_U$ are obtained
by conjugation \cite{AFGNT} with $\sigma$, i.e. $g_i'=\sigma^{-1}g_i\sigma$.

Any $N=2$ heterotic string vacuum
contains two further $U(1)$ vector fields, namely the graviphoton field,
which has no physical scalar partner, and the dilaton-axion field,
denoted by $S$. Thus the Abelian gauge symmetry we 
will be considering in the following is
given by $U(1)_L^2\times U(1)_R^2$.
At special lines in the $(T,U)$ moduli space, additional vector fields
become massless and the $U(1)_L^2$ becomes enlarged to a
non-Abelian gauge symmetry. Specifically, there are four
inequivalent lines in the moduli space where two charged gauge bosons
become massless, namely at $U=T$, $U = \frac{1}{T}$, $U = T-i$ and
$U = \frac{T}{iT + 1}$.  We will, in this paper, only consider the
$T=U\neq 1,\rho$ line for definiteness.  The other 3 lines can be discussed 
in a similar way \cite{CLM2}.

At the critical line $T=U\neq 1,\rho$, the $U(1)_L^2$ is extended to
$SU(2)_{(1)}\times U(1)$. The Weyl reflection $w_1$ of the enhanced
gauge group $SU(2)_{(1)}$ acts as follows on $T$ and $U$, $w_1 : T
\leftrightarrow U$.  Thus, $w_1 = \sigma$ and the Weyl reflection $w_1$
is a target space duality transformation.
Similarly, the Weyl transformations associated with the other 3 
critical lines are also
target space modular transformations
and therefore also elements of the monodromy group. All Weyl reflections
can be expressed in terms of the generators $g_1,g_2$ and $\sigma$.
The four critical lines are fixed under the corresponding
Weyl transformation. The number
of additional massless states at a given critical line/point
agrees with the order of the fixed
point transformation at that critical line/point \cite{CLM}.

The moduli fields $T$ and $U$ can be expressed in terms
of the field theory Higgs fields whose non-vanishing vacuum
expectation values spontaneously break the enlarged gauge
symmetry 
down to $U(1)^2_L$.
The Higgs field of $SU(2)_{(1)}$, for instance, 
is given by $a_1 \propto (T-U)$.

The classical vector couplings are determined by the holomorphic prepotential
which is a homogeneous function of degree two of the fields $X^I$
($I=0,\dots , 3$). It is given by \cite{CAFP,WKLL,AFGNT}
\begin{equation}
F=i{X^1X^2X^3\over X^0}=- STU
\label{classprep}
\end{equation}
where the physical vector fields are defined as $S=i{X^1\over X^0}$,
$T=-i{X^2\over X^0}$, $U=-i{X^3\over X^0}$ and the graviphoton corresponds
to $X^0$. As explained in \cite{CAFP,WKLL}, 
the period vector $(X^I,iF_I)$ ($F_I=
{\partial F\over \partial X^I}$), that 
follows from the prepotential (\ref{classprep}),
does not lead to classical gauge couplings which all become small in the
limit of large $S$.  In order to arrive at a `physical'
period vector \cite{CAFP,WKLL}
one has to perform the
following symplectic transformation $(X^I,iF_I)\rightarrow (P^I,iQ_I)$,
where
$P^1 = i F_1,\; Q_1 = i X^1,\mbox{   and   }P^i = X^i,\;Q_i = F_i\quad
{\mbox {\rm for } }\, i =0,2,3$.
In this new basis the classical period vector takes the form
\begin{equation}
\Omega^T=(1,TU,iT,iU,iSTU,iS,-SU,-ST)
\end{equation}
where $X^0=1$. One sees that 
all electric vector fields $P^I$ depend only on $T$ and $U$, whereas
the magnetic fields $Q_I$ are all proportional to $S$.

The basis $\Omega$ is also well adapted to discuss the action
of the target space duality transformations and, as particular
elements of the target space duality group, of the four inequivalent
Weyl reflections associated with the four critical lines of gauge
symmetry enhancement.  
In general, the field equations of
the $N=2$ supergravity action are invariant under the following
symplectic $Sp(8,{\bf Z})$ transformations, which act on the period
vector $\Omega$ as \cite{CAFP}
\begin{equation}
\pmatrix{P^I\cr i Q_I\cr}\rightarrow \Gamma\pmatrix{
P^I\cr  i Q_I\cr}=\pmatrix{U&Z\cr W&V\cr}\pmatrix{
P^I\cr i Q_I\cr}
\label{symptr}
\end{equation}
where the $4\times 4$ sub-matrices $U,V,W,Z$ have to satisfy the symplectic
constraints
$U^T V - W^T Z = V^T U - Z^T W = 1,\;\;\;
U^T W = W^T U,\;\;\;Z^T V = V^T Z$.
Invariance of the lagrangian implies that $W=Z=0$, $VU^T=1$. In case
that $Z=0$, $W\neq 0$ and hence $VU^T=1$ the action is invariant up
to shifts in the $\theta$-angles.  The non-vanishing matrix $W$
corresponds to a non-trivial one-loop monodromy due to logarithmic
singularities in the prepotential. (see section 3.)
Finally, if $Z\neq 0$ then the electric fields transform into magnetic fields;
these transformations are the non-perturbative monodromies due to
logarithmic singularities induced by monopoles, dyons or other
non-perturbative excitations (see section 4).

As mentioned before,
the classical action is completely invariant under the target space
duality transformations. Thus the classical monodromies have $W,Z=0$.
The classical monodromy
matrix $U$ ($V=U^{T,-1}=U^*$) associated with the Weyl reflection
$w_1=\sigma$ is given by
\be
U_{\sigma} = \left( \begin{array}{cccc}
1&0&0&0\\
0&1&0&0\\
0&0&0&1\\
0&0&1&0\\
\end{array} \right)
\eq
At the classical level the $S$-field is invariant under target space
duality
transformations.

\section{Perturbative results \label{pertsec}}

Loop corrections to the
holomorphic prepotential only occur at one-loop.
Simple power counting arguments imply that the one-loop
correction to the prepotential 
must be independent of the dilaton field $S$. 
Thus the perturbative
prepotential takes the form \cite{WKLL,AFGNT}
\beqa
F &=& F^{(Tree)}(X) + F^{(1-loop)}(X) \nonumber\\
&=& i \frac{X^1 X^2 X^3}{X^0} + (X^0)^2 f(T,U) \nonumber\\
&=& - STU + f(T,U)
\eeqa
Since the target space duality transformations are known to be
a symmetry in each order of perturbation theory, the tree level plus
one-loop effective action must be invariant under these transformations up to
discrete shifts in the various $\theta$
angles 
due to monodromies around semi-classical
singularities in the moduli space where massive string modes become
massless.
The period vector
$\Omega^T = (P^I,i Q_I)$ transforms perturbatively as follows
\begin{equation}
P^I\ \rightarrow\    U{}^I_{\,J}\, P^J,\;
i Q_I\ \rightarrow\    V_I{}^J \,i Q_J\ +\   W_{IJ}\, P^J\ 
\label{symptrans}
\end{equation}
where
\begin{equation}
 V = ( U^{\rm T})^{-1},\quad
 W =  V \Lambda\,,\quad \Lambda=\Lambda^{\rm T}
\end{equation}
and $ U$ belongs to  $SO(2,2,{\bf Z})$.
Classically, $\Lambda=0$, but in the quantum theory, $\Lambda$
is a real symmetric matrix, which should be integer
valued in some basis.

In addition, the effective action is
also invariant, up to discrete shifts in the $\theta$-angles,
under discrete shifts in the $S$-field, $D$: $S\rightarrow S-i$.
Thus, the full perturbative monodromy group  contains 
the following $Sp(8,{\bf Z})$
transformation
\beqa
V_{S} &=& U_S = \left( \begin{array}{cccc}
1&0&0&0\\
0&1&0&0\\
0&0&1&0\\
0&0&0&1\\
\end{array} \right) \nonumber\\
W_S &=& \left( \begin{array}{cccc}
0&1&0&0\\
1&0&0&0\\
0&0&0&1\\
0&0&1&0\\
\end{array} \right),\;\;\; Z_S = 0
\label{sdualpert}
\eeqa
As explained in \cite{WKLL,AFGNT} 
the dilaton is not any longer invariant under
the target space duality transformations at the one-loop level.
Indeed, it can be shown \cite{WKLL} that
(\ref{symptrans}) implies that
\be
S \longrightarrow S + \frac{V_1^{\;J} (F^{(1-loop)}_J -i
\Lambda_{JK} P^K )}{U^{0}_{\;I} P^I}
\label{spert}
\eq
Near the critical lines of classical gauge symmetry enhancement
the one-loop prepotential exhibits
logarithmic singularities and is therefore not a singlevalued function
when transporting the moduli fields around these singular lines.
For example around the singular $SU(2)_{(1)}$ line $T=U \neq 1,\rho$ 
the function $f$
must have the
following form \cite{WKLL,AFGNT,CLM2}
\begin{equation}
f(T,U)={1\over \pi}(T-U)^2\log(T-U)+\Delta(T,U)
\label{fsu21}
\end{equation}
where $\Delta(T,U)$ is finite and single valued at $T=U \neq 1,\rho$.
At the remaining three critical  lines $f(T,U)$ takes an analogous form.

\subsection{Perturbative $SU(2)_{(1)}$ monodromies}

Let us now consider the element $\sigma$
which corresponds to the Weyl reflection in the enhanced $SU(2)_{(1)}$.
Under the 
mirror transformation $\sigma$, 
$T \leftrightarrow U, T-U \rightarrow e^{-i \pi}
(T-U)$, and the
$P$ transform classically and perturbatively as
\be
P^0 \rightarrow P^0,\;\;\;
P^1 \rightarrow P^1,\;\;\;
P^2 \rightarrow P^3,\;\;\;
P^3 \rightarrow P^2
\eq
Using (\ref{fsu21}), 
the 1-loop corrected $Q_2$ and $Q_3$ are computed to be 
\beqa
Q_2 &=& i S U 
- \frac{2i}{\pi} (T-U) \log(T-U) \nonumber\\
&-& \frac{i}{\pi} (T-U) -i \Delta_T \nonumber\\
Q_3 &=& i S T
+ \frac{2i}{\pi} (T-U) \log(T-U) \nonumber\\
&+& \frac{i}{\pi} (T-U)
-i \Delta_U
\label{q2q3}
\eeqa
It follows from (\ref{spert}) that,
under mirror symmetry $T \leftrightarrow U$, 
the dilaton $S$ transforms as 
\be
S \rightarrow S + i
\label{stransf}
\eq
Then, it follows that 
perturbatively
\beqa
\left( \begin{array}{c}
Q_2\\  Q_3\\
\end{array} \right) 
\rightarrow
\left( \begin{array}{c}
Q_3\\  Q_2\\
\end{array} \right) +  
\left( \begin{array}{cc}
1 & -2\\
-2 & 1 \\
\end{array} \right) 
\left( \begin{array}{c}
T\\  U\\
\end{array} \right) 
\label{su2mono}
\eeqa
Thus, 
the section $\Omega$ transforms perturbatively as
$\Omega \rightarrow \Gamma_{\infty}^{w_1} \Omega$, where \cite{AFGNT,CLM2}
\beqa
\Gamma_{\infty}^{w_1} &=& 
\left( \begin{array}{cc}
U_{\sigma} & 0 \\ U_{\sigma} \Lambda & U_{\sigma}\\
\end{array} \right)  \;\;,\;\;
U_{\sigma}=\left( \begin{array}{cc}
I & 0 \\ 0 &  \eta \\
\end{array} \right)  \nonumber\\
\Lambda &=& - 
\left( \begin{array}{cc}
\eta & 0 \\
0 & {\cal C}\\
\end{array} \right) \;\;,\;\;
\eta = \left( \begin{array}{cc}
0& 1 \\
1 & 0\\
\end{array} \right) \nonumber\\
{\cal C} &=& \left( \begin{array}{cc}
2& -1 \\
-1 & 2\\
\end{array} \right) 
\nonumber\\
\label{perttrans}
\eeqa

\subsection{Truncation to the rigid case of Seiberg/Witten \label{secseiwit}}

In order to truncate \cite{CLM2} the perturbative 
$SU(2)_{(1)}$ monodromy matrix $\Gamma_{\infty}^{w_1}$
to the rigid one of Seiberg/Witten \cite{SW1}, we will take the limit
$\kappa^2=\frac{8 \pi}{M^2_{pl}} \rightarrow 0$ and expand
$T = T_0 + \kappa \delta T ,\;
U= T_0 + \kappa \delta U$.
Here we have expanded the moduli fields $T$ and $U$ around the same
vev $T_0 \neq 1,\rho$.  Both $ \delta T$ and $\delta U$ denote 
fluctuating fields of mass dimension one.  We will also freeze in the 
dilaton field to a 
large vev, $S = \langle S \rangle $.  
Then, the $Q_2$ and $Q_3$ given in (\ref{q2q3}) can be expanded 
as
\beqa
Q_2 &=&  i \langle S \rangle T_0 + \kappa \tilde{Q}_2 \;\;\;,\;\;\;
Q_3 = i \langle S \rangle T_0 + \kappa \tilde{Q}_3 \nonumber\\
\tilde{Q}_2 &=& i \langle S \rangle \delta U 
- \frac{2i}{\pi} (\delta T- \delta U) 
\log\kappa ( \delta T - \delta U) \nonumber\\
&-& \frac{i}{\pi} (\delta T- \delta U) 
- i \Delta_T(\delta T,\delta U) \nonumber\\
\tilde{Q}_3 &=& i \langle S \rangle \delta T
+ \frac{2i}{\pi} (\delta T- \delta U) 
\log\kappa ( \delta T - \delta U) \nonumber\\
&+& \frac{i}{\pi} (\delta T- \delta U) 
- i \Delta_U(\delta T,\delta U)
\label{truncq2q3}
\eeqa
Next, one has to specify how mirror symmetry is to act on 
the vev's $T_0$ and $\langle S \rangle$ as well as on $\delta T$ and
$\delta U$.  We will take that under mirror symmetry
\beqa
T_0 \rightarrow T_0 \;\;,\;\; \delta T \leftrightarrow \delta U \;\;,\;\;
\langle S \rangle \rightarrow
\langle S \rangle 
\label{tSTU}
\eeqa
Note that we have taken $\langle S \rangle$ to be invariant under mirror
symmetry.  This is an important difference to (\ref{stransf}).
Using (\ref{tSTU}) and that $\delta T - \delta U \rightarrow
e^{-i \pi} (\delta T - \delta U)$, it follows that the truncated
quantities $\tilde{Q}_2$ and $\tilde{Q}_3$ transform as 
follows under mirror symmetry
\beqa
\left( \begin{array}{c}
\tilde{Q}_2\\  \tilde{Q}_3\\
\end{array} \right) 
\rightarrow
\left( \begin{array}{c}
\tilde{Q}_3\\  \tilde{Q}_2\\
\end{array} \right) +  
\left( \begin{array}{cc}
2 & -2\\
-2 & 2 \\
\end{array} \right) 
\left( \begin{array}{c}
\delta T\\ \delta  U\\
\end{array} \right) 
\label{truncsu2mono}
\eeqa
Defing a truncated section $\tilde{\Omega}^T= (\tilde{P}^2,\tilde{P}^3,
i\tilde{Q}_2,i\tilde{Q}_3)=(i \delta T,i \delta U,
i\tilde{Q}_2,i\tilde{Q}_3)$, it follows that $\tilde{\Omega}$ transforms
as $\tilde{\Omega} \rightarrow \tilde{\Gamma}_{\infty}^{w_1} \tilde{\Omega}$
under mirror symmetry (\ref{tSTU}), where
\beqa
\tilde{\Gamma}_{\infty}^{w_1} &=& 
\left( \begin{array}{cc}
\tilde{U} & 0 \\ \tilde{U} \tilde{\Lambda} & \tilde{U}\\
\end{array} \right)  \;\;,\;\;
\tilde{U}= \left( \begin{array}{cc}
0& 1 \\
1 & 0\\
\end{array} \right) \nonumber\\
\tilde{\Lambda} &=& \left( \begin{array}{cc}
-2& 2 \\
2 & -2\\
\end{array} \right) 
\label{truncmonotrans}
\eeqa
Note that, because of the invariance of $\langle S \rangle$ under mirror
symmetry, $\tilde{\Lambda} \neq - {\cal C}$, contrary to what one
would have gotten by performing a naive truncation of (\ref{perttrans})
consisting in keeping only rows and columns associated with 
$(P^2,P^3,iQ_2,iQ_3)$.

Finally, in order to compare the truncated $SU(2)$ monodromy 
(\ref{truncmonotrans}) with the perturbative $SU(2)$ monodromy
of Seiberg/Witten \cite{SW1}, one has to perform a change of basis from 
moduli fields to Higgs fields, as follows 
\beqa
\left( \begin{array}{c}
a \\
a_D\\
\end{array} \right) 
&=&
M \tilde{\Omega} \;\;,\;\;
M = 
\left( \begin{array}{cc}
m &  \\
 & m^* \\
\end{array} \right)  \nonumber\\
m &=& \frac{\gamma}{\sqrt{2}}
\left( \begin{array}{cc}
1& -1 \\
1& 1\\
\end{array} \right) 
\label{higgsbasis}
\eeqa
where $\gamma$ denotes a constant to be fixed below.  Then, the 
perturbative $SU(2)$ monodromy in the Higgs basis is given by
\beqa
\tilde{\Gamma}^{Higgs}_{\infty} &=& M \tilde{\Gamma}_{\infty}^{w_1} M^{-1}
\nonumber\\
&=&
\left( \begin{array}{cc}
m \tilde{U} m^{-1}  &   0  \\
m^* \tilde{U} \tilde{\Lambda} m^{-1} &
m^* \tilde{U} m^T \\
\end{array} \right) 
\eeqa
which is computed to be 
\beqa
\tilde{\Gamma}^{Higgs,w_1}_{\infty} 
=
\left( \begin{array}{cccc}
-1 & & & \\
& 1 & & \\
\frac{4}{\gamma^2} & 0 & -1 & 0 \\
& & & 1\\
\end{array} \right) 
\label{higgsmono}
\eeqa
The monodromy matrix (\ref{higgsmono}) correctly reproduces
the
perturbative $SU(2)$ monodromy of Seiberg/Witten \cite{SW1} for
$\gamma^2 = 2$, whereas comparision with the perturbative $SU(2)$
monodromy of Klemm et al \cite{KLT} gives that $\gamma^2 = 1$.

\section{Non perturbative monodromies}

In order to obtain some information about non-perturbative
monodromies in $N=2$ heterotic string compactifications, we will follow
Seiberg/Witten's strategy in the rigid case \cite{SW1} 
and we will first try to decompose the
perturbative monodromy
matricx $\Gamma_{\infty}^{w_1}$ into
$\Gamma_{\infty}^{w_1} = \Gamma_M^{w_1} 
\Gamma_D^{w_1}$ with $\Gamma_M^{w_1}$ ($\Gamma_D^{w_1}$)
possessing monopole (dyonic) like fixed points. Thus,
the critical line $T=U$ of classical gauge symmetry enhancement
will split into two non-perturbative singular lines where
magnetic monopoles or dyons respectively become massless. 
Similar considerations can be made for any of the other singular
lines \cite{CLM2}.

\subsection{Non perturbative monodromies for $SU(2)_{(1)}$ 
\label{su21monopol}}

We will assume that

\begin{enumerate}
\item
$\Gamma_{\infty}^{w_1}$ is to be decomposed into precisely two factors,
namely
\be
\Gamma_{\infty}^{w_1} = \Gamma_{M}^{w_1} \Gamma_{D}^{w_1}
\label{gmgd}
\eq
\item
$\Gamma_{M}^{w_1}$ and therefore $\Gamma_{D}^{w_1}$ must be symplectic.
\item
$\Gamma_{M}^{w_1}$ ($\Gamma_{D}^{w_1}$)
must have a monopole (dyon) like fixed point.
\item
$\Gamma_{M}^{w_1}$ and $\Gamma_{D}^{w_1}$ 
should be conjugated, 
as it is the case in the rigid theory.
\item
$\Gamma_{\infty}^{w_1}$ has a peculiar block structure in that
the indices $j=0,1$ of the section $(P_j, i Q_j)$ are never
mixed with the indices $j=2,3$.
We will assume that $\Gamma_{M}^{w_1}$ and $\Gamma_{D}^{w_1}$ also have
this structure. 
\item
We 
will take $\Gamma_{M}^{w_1}$ to be the identity matrix 
on its diagonal.
The existence of a basis
where the non--perturbative monodromies have this special form
will be aposteriori justified by the fact that it leads to
a consistent
truncation to the rigid case.

\end{enumerate}

Putting all these things together yields \cite{CLM2} the following 8$\times$8
non--perturbative
monodromy matrix $\Gamma_M^{w_1}$ 
that depends on four parameters $x,y,v$ and $p$ and that
consistently describes the splitting of
the $T=U$ line
\be
\Gamma_{M}^{w_1} =
\left [\begin {array}{cccccccc} 1&0&0&0&0&0&0&0\\\noalign{\medskip}0&
1&0&0&0&0&0&0\\\noalign{\medskip}0&0&1&0&0&0&-2/3&2/3
\\\noalign{\medskip}0&0&0&1&0&0&2/3&-2/3\\\noalign{\medskip}x&y&0&0&
1&0&0&0\\\noalign{\medskip}y&v&0&0&0&1&0&0\\\noalign{\medskip}0&0&p&p
&0&0&1&0\\\noalign{\medskip}0&0&p&p&0&0&0&1\end {array}\right ]
\label{monopolew1}
\eq
$\Gamma_D^{w_1}$ then immediately follows from (\ref{gmgd}).

The associated fixed points have the form
\be
\left(N,-M \right) = 
\left( 0 , 0 , N^2, -N^2, 0, 0, 0, 0 \right)
\eq
for the monopole and
\be
\left(N, -M \right) = 
\left(0, 0,  N^2, -N^2, 0, 0, \frac{3}{2}
N^2, -\frac {3}{2} N^2 \right)
\eq
for the dyon.

Note that demanding the monopole matrix to be conjugated
to the dyonic monodromy matrix leads to the requirement
$p \neq 0$ \cite{CLM2}.

\subsection{Truncating the $SU(2)_{(1)}$ monopole monodromy
to the rigid case}

The monopole monodromy matrix $\Gamma_M^{w_1}$
depends on four yet undetermined parameters, namely
$x,v,y$ and $p\neq 0$.  One way to determine their values
is to demand that, upon truncation of (\ref{monopolew1})
to the rigid case, one recovers the rigid non-perturbative
monodromies of Seiberg/Witten.

Consider the $4 \times 4$ monopole subblock 
which acts on $(P^2,P^3,iQ_2,iQ_3)$
\beqa
\Gamma_{M23}^{w_1}=
\left [\begin {array}{cccc} 
1&0&-2 \alpha &2 \alpha
\\\noalign{\medskip}0&1&2 \alpha &-2 \alpha 
\\\noalign{\medskip}
p&p&1&0\\\noalign{\medskip}p&p&0&1\end {array}\right ] 
\eeqa
where
$\alpha = \frac{1}{3}, p \neq 0$.
Rotating it into the Higgs basis gives that
\beqa
\Gamma_{M}^{Higgs,w_1} &=& M
\Gamma_{M23}^{w_1} M^{-1} \nonumber\\
&=& 
\left [\begin {array}{cccc} 
1&0&- 4\alpha \gamma^2&0
\\\noalign{\medskip}0&1&0&0
\\\noalign{\medskip}
0&0&1&0\\\noalign{\medskip}0&\frac{2p}{\gamma^2}&0&1\end {array}\right ]
\label{locmonohiggs}
\eeqa
where $M$ is given in equation (\ref{higgsbasis}).
In the rigid case, on the other hand, one expects to find 
for the rigid monopole monodromy matrix in the Higgs basis that
\beqa
\tilde{\Gamma}_{M}^{Higgs,w_1}=
\left [\begin {array}{cccc} 
1&0&- 4\tilde{\alpha}\gamma^2&0
\\\noalign{\medskip}0&1&0&0
\\\noalign{\medskip}
0&0&1&0\\\noalign{\medskip}0&\frac{2\tilde{p}}{\gamma^2}
&0&1\end {array}\right ]
\label{rigidmono}
\eeqa
where $\tilde{\alpha}=\frac{1}{4}, \tilde{p} =0$.
The first and third lines of (\ref{rigidmono}) are 
nothing but the monodromy matrix
for one $SU(2)$ monopole ($\gamma^2 = 2$ in the conventions of 
Seiberg/Witten \cite{SW1}, 
and $\gamma^2 = 1 $ in the conventions of Klemm et al \cite{KLT}).
Thus, truncating the monopole monodromy matrix (\ref{monopolew1})
to the rigid case appears to produce jumps in the parameters $p \rightarrow
\tilde{p}=0$ and $\alpha \rightarrow \tilde{\alpha}$ as given above.
These jumps are due to the freezing in of the dilaton (see eq.(\ref{tSTU}))
when taking the flat limit.
In \cite{CLM2} we presented a field theoretical explanation
for the jumps occuring in the parameters $p$ and $\alpha$ when 
taking the rigid limit
described in section \ref{secseiwit}.  
This explanation also determines, as a bonus,
the values of the parameters $v, y$ and $p$, namely
$y = \frac{8}{3}$, $p = \frac{4}{3}$ and $v=0$.
Moreover, one can show that, in order
to decouple the four $U(1)$'s at the non-perturbative level, one
has to have $x=v$ and consequently $x=0$.  

\section{Comparision with type II}

The monopole monodromy matrix $\Gamma_M^{w_1}$ given in (\ref{monopolew1}),
which was obtained by decomposing $\Gamma_{\infty}^{w_1}$, is not an
integer valued matrix.  
Recall that 
$\Gamma_{\infty}^{w_1}$ doesn't leave the dilaton $S$ invariant, but
rather induces the shift $S \rightarrow S+i$.  Thus, consider \cite{AntoPartou}
combining $\Gamma_{\infty}^{w_1}$ with a compensating shift $S \rightarrow
S-i$ for the dilaton, which is generated by $\Gamma_S$ given in 
(\ref{sdualpert}).  Then, under $\hat{\Gamma}_{\infty}^{w_1} = \Gamma_S
\Gamma_{\infty}^{w_1}$, $T \leftrightarrow U$ whereas $S \rightarrow S$.
$\hat{\Gamma}_{\infty}^{w_1}$ is given by 
\beqa
\hat{\Gamma}_{\infty}^{w_1} &=& 
\left( \begin{array}{cc}
U_{\sigma} & 0 \\ X & U_{\sigma}\\
\end{array} \right)  \;\;,\;\;
X =  
\left( \begin{array}{cc}
0 & 0 \\
0 & {\tilde U} {\tilde \Lambda}\\
\end{array} \right) 
\eeqa
Comparing with (\ref{truncmonotrans}) shows that the non-trivial submatrix of
$\hat{\Gamma}_{\infty}^{w_1}$, which acts only on the rows and columns
associated with $(P^2,P^3,iQ_2,iQ_3)$, turns in the Higgs basis 
(\ref{higgsbasis})
precisely into the rigid perturbative $SU(2)$ monodromy matrix of 
Seiberg/Witten.  Thus, $\hat{\Gamma}_{\infty}^{w_1}$ can be straightforwardly
decomposed into $\hat{\Gamma}_{\infty}^{w_1}=\hat{\Gamma}_{M}^{w_1}
\hat{\Gamma}_{D}^{w_1}$.  
The integer valued monopole matrix 
$\hat{\Gamma}_{M}^{w_1}$, for instance,
reads
\beqa
\hat{\Gamma}_{M}^{w_1} &=&
\left [\begin {array}{cccccccc} 1&0&0&0&0&0&0&0\\\noalign{\medskip}0&
1&0&0&0&0&0&0\\\noalign{\medskip}0&0&1&0&0&0&-1&1
\\\noalign{\medskip}0&0&0&1&0&0&1&-1\\\noalign{\medskip}0&0&0&0&
1&0&0&0\\\noalign{\medskip}0&0&0&0&0&1&0&0\\\noalign{\medskip}0&0&0&0
&0&0&1&0\\\noalign{\medskip}0&0&0&0&0&0&0&1\end {array}\right ]
\eeqa
The decomposition of $\hat{\Gamma}_{\infty}^{w_1}$ leads
to integer valued non perturbative monodromy matrices 
$\hat{\Gamma}_M^{w_1}$ and $\hat{\Gamma}_D^{w_1}$, which are in 
agreement
with the ones computed on the type II Calabi-Yau side 
\cite{AntoPartou,KKLMV}.

\end{document}